\documentclass[prd,showpacs]{revtex4}
\usepackage{graphics}
\def \beq{\begin{equation}}
\def \eeq{\end{equation}}
\def \beqa{\begin{eqnarray}}
\def \eeqa{\end{eqnarray}}

\def \alphas{\alpha_{\scriptscriptstyle S}}
\begin{document}

\title{Susceptibilities and screening masses in two flavor QCD}
\author{Rajiv V.\ Gavai}
\email{gavai@tifr.res.in}
\author{Sourendu Gupta}
\email{sgupta@tifr.res.in}
\author{Pushan Majumdar}
\email{pushan@theory.tifr.res.in}
\affiliation{Department of Theoretical Physics, Tata Institute of Fundamental
         Research,\\ Homi Bhabha Road, Mumbai 400005, India.}

\begin{abstract}
We studied QCD with two flavors of dynamical staggered quarks at
finite temperature, with a bare sea quark mass of about 17 MeV. We
report investigations of baryon, isospin, charge and strangeness
susceptibilities, as well as screening masses obtained from correlators
of local and one-link separated meson operators. These were studied as
functions of valence quark mass at several temperatures. Our results
for susceptibilities deviate significantly from ideal gas values, and
even more from the weak coupling series.  We also report the first
measurement of off-diagonal quark number susceptibilities below the
transition temperature, $T_c$, where they are the main contribution to
charge fluctuations. We present evidence for a close connection between
the susceptibilities and the screening masses.
\end{abstract}
\pacs{11.15.Ha, 12.38.Mh}
\preprint{TIFR/TH/01-38, hep-ph/0110032}
\maketitle

\section{Introduction}

Experiments at the Brookhaven RHIC are now seeking to establish the
formation of quark-gluon plasma. Proposals for experimental signatures of
the plasma have to take into account its physical nature. Such information
is most reliably obtained in lattice simulations of the hot phase of
QCD. At present the phase transition temperature, $T_c$, is fairly well
known \cite{tc,tc2}, and reasonably reliable information on the equation
of state has also been obtained \cite{eos}. Screening and fluctuations
of conserved charges are some of the other relevant properties of the
plasma. We report extensive new lattice results for these.

We address two different physics questions in this paper.  We are
primarily interested in the question of baryon number and electric charge
fluctuations in the plasma. These are quantities of direct experimental
relevance, as has been realized recently \cite{koch,chi0s}. We construct
them from measurements of quark number susceptibilities \cite{qsdef}. In
addition, we study screening masses in the plasma through the study of
spatial correlators, paying special attention to their tensor structures.
This gives a new relation between the susceptibilities and screening
masses which are verified by our lattice measurements.  In particular,
we explain why non-perturbative phenomena in the latter are closely
connected with deviations from perturbation theory in the former.

Temperatures just above $T_c$ are likely to be the most relevant region
for applications in heavy ion physics and cosmology.  Lattice measurements
of the equation of state deviate significantly from the usual high
temperature perturbative expansion.  This stimulated many attempts
to resum the weak coupling expansion, with varying degrees of success
\cite{resum}.  Many of these techniques would have definite predictions
for the susceptibilities that we measure. Since our measurements also
deviate from the weak coupling expansion, they stand as yet another
invitation to resum the weak coupling series.

While the world contains six flavors of quarks, it suffices to consider
two light (u and d) and a moderately heavy (s) quark for the physics
of the QCD phase transition.  Starting from a quenched theory where
all quark loop effects are turned off, one can envisage successive
approximations by which dynamical quarks are switched on in the sequence
of their masses. Such an incremental strategy is actually forced by
the cost of lattice simulations of full QCD. Earlier we simulated QCD
in the quenched approximation, and reported our measurements of various
quark susceptibilities \cite{qsus}.  In the simulations we report here,
we go to the next level of approximation by using dynamical u and d
quarks with a bare mass of about 17 MeV.  Our results indicate that
at temperatures above $T_c$ the change due to the inclusion of light
quarks is small. Unquenching the heavier strange quark may thus be
correspondingly less important. Even in the cold phase below $T_c$,
taking still lighter u and d quarks may be more crucial in future than
including dynamical strange quarks.

The plan of this paper is the following--- technical material on
the simulations, including the lattice scales and details of the data
taking procedure, is presented in the next section. The definitions and
notation, and details of measurements of the screening masses and quark
number susceptibilities are given in the following two sections, in that
order. The discussion and summary in the final section is designed to
facilitate use by those who are mainly interested in applications of
our results or wish to have an overview of future directions in lattice
measurements of susceptibilities.

\section{The simulations and scales}

\subsection{Run parameters}

\begin{table}[!htbp]
  \begin{center}\begin{tabular}{|c|c|c|c|c|}  \hline
  $T/T_c$ & $ma$ & $\beta$ & size & statistics \\
  \hline
  0.00 & 0.025  &5.2875 & $12^4$ &   81 \\
  1.00 & 0.025  &5.2875 & $4\times12^3$ & 249 \\
       &        &       & $4\times16^3$ &  42 \\
  1.25 & 0.02   & 5.35  & $4\times12^3$ &   50 \\
  1.50 & 0.0125 & 5.420 & $4\times12^3$  &   50 \\
  1.50 & 0.0167 & 5.429 & $4\times8^3$  &   50 \\
       &        &       & $4\times12^3$         &    50 \\
       &        &       & $4\times16^3$         &    20 \\
       &        &       & $4\times16\times10^2$ &  2001 \\
       &        &       & $4\times16\times10^2$ &  $124^*$ \\
       &        &       & $4\times24\times10^2$ &  1500 \\
  2.00 & 0.0125 & 5.540 & $4\times12^3$ &    50 \\
       &        &       & $4\times16\times10^2$ &  2021 \\
  3.00 & 0.0083 & 5.675 & $4\times12^3$ &    50 \\
       &        &       & $4\times16^3$ &   1139 \\
  \hline
  \end{tabular}\end{center}
  \caption[dummy]{Details of $N_f=2$ runs. The runs were made with a
     trajectory length of 1 MD time unit, a time step of $0.01$ and a
     conjugate gradient stopping criterion of $10^{-5}\sqrt V$. The
     starred run was performed with half the MD time step but the same
     trajectory length. }
\label{tb.runs}\end{table}

\begin{table}[!bhtp]
  \begin{center}\begin{tabular}{|c|c|c|c|c|}  \hline
  $T/T_c$ & $ma$ & $\beta$ & size & statistics \\
  \hline
  1.5  & 0.0167 & 5.8941 & $4\times24\times10^2$ &  800 \\
  2.0  & 0.0125 & 6.0625 & $4\times16\times10^2$ & 6500 \\
       &        &        & $4\times24\times10^2$ &  850 \\
  3.0  & 0.0083 & 6.3375 & $4\times16\times14^2$ & 7500 \\
  \hline
  \end{tabular}\end{center}
  \caption[dummy]{Details of new quenched runs. These runs supplement
     earlier ones discussed in \protect\cite{qsus}.}
\label{tb.qruns}\end{table}

We have simulated QCD with two flavors of light dynamical staggered
quarks ($N_f=2$) with the R-algorithm \cite{hybridr} on $N_t\times N_x^2
\times N_z$ lattices with $N_t=4$. For investigations of quark number
susceptibilities, we chose spatially symmetric lattices with $N_x=N_z$.
On the other hand, screening correlators were easier to measure on
elongated lattices with $N_z>N_x$.  Most of our simulations consisted
of generating dynamical configurations with two flavors ($N_f=2$) of
dynamical staggered quarks with the sea quark mass, $m$, held fixed in
physical units at $m/T_c=0.1$ as the temperature was varied.  A small
set of runs with $m/T_c=0.075$ was performed to check the magnitude
of sea quark mass effects. The run parameters and statistics collected
are shown in Table \ref{tb.runs}.  We emphasize that holding the quark
mass fixed is important for physics applications, whereas the older
measurements fixed $m/T$ instead.

After a few tuning runs which reproduced the results of previous studies
\cite{hybridr,nf2}, we chose to run with parameters similar to those used
in previous works with two flavors of staggered quarks, {\sl i.e.\/},
trajectory length of 1 molecular dynamics (MD) time unit, integrated in
100 steps of 0.01 MD time units each, and a conjugate gradient stopping
criterion of $10^{-5}\sqrt V$ on the modulus of the remainder vector ($V =
N_x^2 N_z N_t$).  It should be noted that very similar run parameters are
often used in $N_f=4$ Hybrid Monte Carlo (HMC) runs, where a Metropolis
choice ensures that the correct weight for a configuration is obtained
even for a finite number of time steps. For $N_f=2$ a bias-free estimate
of the weight is not guaranteed. In order to estimate systematic errors
from this source we also made a test run with half the step size but
for the same trajectory length.

Part of the purpose of this work is a systematic comparison of screening
masses with different numbers of dynamical quarks. Details of our $N_f=4$
simulations are available elsewhere \cite{b1,dimred}.  Similar details
of runs in quenched QCD have been presented in \cite{qsus}.  However,
we also generated a few sets of new quenched configurations.  Details of
the algorithm and data taking remain as in \cite{qsus}. The run parameters
and statistics for these quenched runs are listed in Table \ref{tb.qruns}.

\begin{figure}[!tbh]
   \includegraphics{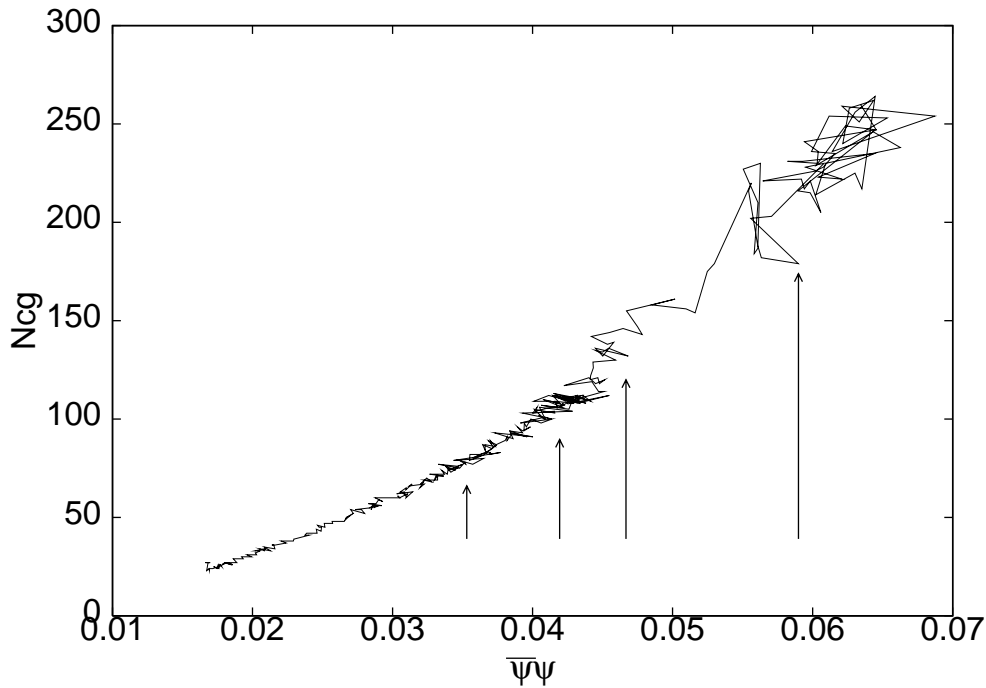}
   \caption{A diagnostic for thermalisation. The points at which the
      trajectory length was changed are marked by vertical arrows.}
\label{fg.thermal}\end{figure}

\subsection{Setting the scale}

Relative temperature scales, {\sl i.e.\/}, $T/T_c$, can be deduced from
previous estimates of critical couplings for various values of $N_t$ at
the quark masses of interest to us \cite{nf2}.  Thus, the temperature
of a run with $N_t=4$ at $\beta_c(N'_t)$ is $T/T_c=N'_t/4$. Since
$T_c$ is known to be $167\pm17$ MeV in physical units \cite{tc2}, this
allows us to deduce the value of lattice spacing, $a$, in physical units
using the relation $T=1/(N_t a)$. For those values of $T$ which cannot
be obtained by simulations on lattices with different $N_t$ we use
the two-loop beta-function of QCD to deduce the lattice spacing, and
hence the temperature. A recent global analysis of data assures us that
this is possible \cite{tc2}. The values of the inverse lattice spacings
appropriate to our measurements are given in Table \ref{tb.runs}. The sea
quark masses that we use are $m=17\pm2$ MeV ($m/T_c=0.1$) and $12\pm1$
MeV ($m/T_c=0.075$).

Another way of setting the lattice spacing is to use previous lattice
measurements of the $\rho$ meson mass with the same quark mass and
coupling \cite{rho}. This gives us a lattice spacing $1/a=580\pm20$
MeV at $T_c$, and hence the dynamical sea quark mass comes out to be
$m=14.5\pm0.6$ MeV ($m/T_c=0.1$) and $m=10.9\pm0.4$ ($m/T_c=0.075$). This
consistency between the two estimates is, of course, not unexpected,
since it is well known that $T_c/m_\rho$ is reasonably independent of
lattice spacing \cite{nf2}.

The strange quark mass can be chosen in two different ways. One is
to take the strange quark mass to be 75-170 MeV \cite{pdg}. This gives
$m_s/T_c=0.41$--$1.0$ \footnote{In quenched QCD, where $T_c\approx280\pm10$
MeV, $m_s/T_c=0.25$--$0.60$.}.  Alternatively, we can use the ratio
$m_s/m_d=17$--$25$ \cite{pdg} to get $m_s/T_c=1.7$--$2.5$. Clearly,
in the limit when the light quark masses are more realistic, the
two procedures should give similar values. However, the heavier the
strange quark mass is, the more strongly will it be subject to lattice
artifacts. In view of this, we have chosen to work with $m_s/T_c=0.75$,
corresponding to $m_s=125$ MeV. A realistic ratio of the strange and light
quark masses will thus need future simulations with lighter u and d quark
masses. Furthermore, the fact that the strange quark is reasonably heavy
in lattice units indicates that the unquenching effects on it would be
small, at least near $T_c$.

\subsection{Thermalisation and autocorrelations}

Due to the fact that each configuration takes a long time to generate,
a major issue in any Fermion simulation is to identify when a run
has thermalised. We monitored thermalisation through measurements
of plaquettes, Wilson line, the chiral condensate and the number of
conjugate gradient iterations needed. Since plaquette measurements
are not very noisy, these were a good estimator of the approach to
equilibrium. At temperatures above $T_c$, the difference of purely
spatial and mixed time-space plaquettes is nonzero, and it turns
out to develop late during equilibration. This was one of the best
criteria we observed for closeness of approach to equilibrium. However,
even more efficient was a scatter plot of the stochastic estimator of
the condensate ($\overline\psi\psi$) against the number of conjugate
gradient iterations ($N_{cg}$).  As shown in Figure \ref{fg.thermal},
a very definite directional movement occurs away from equilibrium,
but as equilibrium is approached the movement becomes a random walk
within a well defined area in the plane of $\overline\psi\psi$ and
$N_{cg}$. Monitoring this scatter plot thus offered a possibility of
tuning the algorithm to achieve fast thermalisation.

We did this by starting each run from an ordered configuration with small
trajectory lengths ($t\simeq0.05$ MD units). As the simulation proceeded,
the plot of $\overline\psi\psi$ against $N_{cg}$ began to slow. When
this happened we increased the trajectory length in small steps until
it reached unity or resulted in further motion. A post-facto check of
the tuning was obtained after subsequent runs verified that the system
continued to perform a random walk without directed movement. This method
was developed in our earlier $N_f=4$ HMC runs and worked well also for
these $N_f=2$ HMD runs. Using this technique we often managed to achieve
thermalisation in 20--30 units of MD time.

A second issue in the control of computer time is the matter of
how often measurements of an observable can be made so that two
successive measurements are effectively decorrelated.  We have estimated
autocorrelation times from a variety of gauge and fermion observables
and found them to be small.  Local operators such as plaquettes seem to
be essentially decorrelated in about 2 trajectories. Even long distance
observables, such as the pion screening correlator at the longest
possible distance, are decorrelated in about 2--3 trajectories.
In view of these results, we have taken data for screening masses on
every second trajectory, and for susceptibilities every tenth trajectory.
This is justified, {\sl a posteriori\/}, by the fact that none of the
screening masses is found to be very small.

\section{Screening correlators and masses}

\begin{table}[htbp]\begin{center}
  \begin{tabular}{|c|c|c|c|c|c|c|}  \hline
  $GRF$ & $\check G$ & $D_4^h$ & Operator & spin$\otimes$flavor & Particle\\
  \hline
  ${\bf 1}^{++}$ & ${\bf 1}_0$ & $A_1^+$ 
      & $\bar\chi(r)\chi(r)$ & $1\otimes1$ & $\pi$, $f_0$ ($\sigma$) \\

  ${\bf 1}^{+-}$ & ${\bf 1}_0$ & $A_1^+$ 
      & $\eta_z(r)\zeta_z(r)\bar\chi(r)\chi(r)$ 
      & $\gamma_z\gamma_5\otimes\gamma_z\gamma_5$, $\gamma_5\otimes\gamma_5$
      & $\pi$ \\

  ${{\bf 3}''''}^{+-}$ & ${\bf 1}_4$ & $A_1^+$
       & $\epsilon(r)\eta_t(r)\zeta_t(r)\bar\chi(r)\chi(r)$
       & $\gamma_i\gamma_5\otimes\gamma_i\gamma_5$,
         $\gamma_i\gamma_z\otimes\gamma_i\gamma_z$
       & $\rho$, $a_1$ \\

       & ${\bf 2}_4$ & $A_1^+$
       & $\epsilon(r)[\eta_x(r)\zeta_x(r)+\eta_y(r)\zeta_y(r)]
                    \bar\chi(r)\chi(r)$ & & \\

       & & $B_1^+$
       & $\epsilon(r)[\eta_x(r)\zeta_x(r)-\eta_y(r)\zeta_y(r)]
                    \bar\chi(r)\chi(r)$ & & \\

  ${{\bf 3}''''}^{++}$ & ${\bf 1}_4$ & $A_1^+$
       & $\epsilon(r)\eta_z(r)\zeta_z(r)\eta_t(r)\zeta_t(r)
                    \bar\chi(r)\chi(r)$
       & $\gamma_i\gamma_k\otimes\gamma_j\gamma_k$,
         $\gamma_i\otimes\gamma_i$
       & $\rho$, $b_1$ \\

       & ${\bf 2}_4$ & $A_1^+$
       & $\epsilon(r)\eta_z(r)\zeta_z(r)[\eta_x(r)\zeta_x(r)+
                    \eta_y(r)\zeta_y(r)]\bar\chi(r)\chi(r)$ & & \\

       & & $B_1^+$
       & $\epsilon(r)\eta_t(r)\zeta_t(r)[\eta_x(r)\zeta_x(r)-
                    \eta_y(r)\zeta_y(r)]\bar\chi(r)\chi(r)$ & & \\

  ${\bf 3}^{-+}$ & ${\bf 1}_6$ & $A_2^-$ 
      & $\eta_t(r)\bar\chi(r)D_t\chi(r)$ 
      & $\gamma_i\otimes1$, $\gamma_i\gamma_k\otimes\gamma_z\gamma_5$
      & $\omega$, $b_1$ \\

      & ${\bf 2}_2$ & $E^-$   
      & $\eta_{x,y}(r)\bar\chi(r)D_{x,y}\chi(r)$ & & \\

  ${\bf 3}^{--}$ & ${\bf 1}_6$ & $A_2^-$
      & $\eta_z(r)\zeta_z(r)\eta_t(r)\bar\chi(r)D_t\chi(r)$ 
      & $\gamma_i\gamma_z\otimes\gamma_z$, $\gamma_i\gamma_5\otimes\gamma_5$
      & $\rho$, $a_1$ \\

      & ${\bf 2}_2$ & $E^-$
      & $\eta_z(r)\zeta_z(r)\eta_{x,y}(r)\bar\chi(r)D_{x,y}\chi(r)$ & & \\
  ${{\bf 3}''}^{--}$ & ${\bf 1}_2$ & $A_2^-$
      & $\epsilon(r)\zeta_t(r)\bar\chi(r)D_t\chi(r)$
      & $\gamma_5\otimes\gamma_i\gamma_5$, $\gamma_z\otimes\gamma_i\gamma_z$
      & $\pi$ \\

      & ${\bf 2}_0$ & $E^-$
      & $\epsilon(r)\zeta_{x,y}(r)\bar\chi(r)D_{x,y}\chi(r)$ & & \\

  ${{\bf 3}''}^{-+}$ & ${\bf 1}_2$ & $A_2^-$
      & $\eta_z(r)\zeta_z(r)\epsilon(r)\zeta_t(r)\bar\chi(r)D_t\chi(r)$
      & $\gamma_z\gamma_5\otimes\gamma_j\gamma_k$, $1\otimes\gamma_i$
      & $\pi$, $a_0$ \\

      & ${\bf 2}_0$ & $E^-$
      & $\eta_z(r)\zeta_z(r)\epsilon(r)\zeta_{x,y}(r)\bar\chi(r)D_{x,y}\chi(r)$
      & & \\
  \hline
  \end{tabular}\end{center}
  \caption[dummy]{Representations of local and one-link separated staggered
     mesons that were used in this study. The table is extracted from
     \protect\cite{d4h}. Here
     $D_\mu\phi({\bf x})=\phi({\bf x}+\hat\mu)+\phi({\bf x}-\hat\mu)$. We
     have chosen $\eta_x=(-1)^{y+z+t}$, $\eta_y=(-1)^{z+t}$,
     $\eta_z=(-1)^t$, $\eta_t=1$, and $\zeta_x=1$, $\zeta_y=(-1)^x$,
     $\zeta_z=(-1)^{x+y}$, $\zeta_t=(-1)^{x+y+z}$. The spin/flavor and
     particle assignments at $T=0$ are standard. $GRF$ is the symmetry
     group of the $T=0$ staggered fermion transfer matrix and $\check G$
     that at $T>0$.}
\label{tb.meson1}\end{table}

\subsection{Symmetries and transfer matrices}

Screening involves the transfer matrix in a spatial direction of the
Euclidean finite temperature lattice. Thus, screening correlators and
masses are classified according to the symmetries of this transfer matrix,
Screening masses are obtained from the exponential decay of the screening
correlators
\beq
   C(z) = \left\langle\frac1{N_x^2N_t}\sum_{x,y,t}
                 M(0)M^\dag(x,y,z,t)\right\rangle.
\label{corr}\eeq
The operators $M(x,y,z,t)$ are made from combinations of quark and gluon
fields which transform according to an irreducible representation (irrep)
of the symmetry group, and the angular brackets denote averaging over
the correct thermal distribution of fields.  We shall also have occasion
to use the meson susceptibilities
\beq
   \chi = \sum_z C(z) =
      \left\langle\frac{N_z}V\sum_{x,y,z,t}M(0)M^\dag(x,y,z,t)\right\rangle.
\label{msus}\eeq
Detailed discussions of the isometries of the lattice and irreps of the
transfer matrix can be found for gluon operators in \cite{bernd,saumen}
and staggered quark operators in \cite{d4h}.  Here we only mention the
features used in this work.

For operators made entirely out of gluon fields, ``glueball'' operators
in the usual shorthand, the relevant symmetry is that of a slice of the
lattice.  For screening correlators such a slice includes two spatial
directions and the Euclidean time direction. Since rotations of the
spatial directions into Euclidean time are disallowed, the full cubic
group, $O_h$, breaks down to the dihedral group, $D_4^h$.  For bilinear
operators constructed from staggered quark fields, the ``meson''
operators, the symmetries of the transfer matrix are more complicated
due to mixing of spin and flavor components and staggering of
the quark fields.  However, this transfer matrix also carries
representations of $D_4^h$ as shown in \cite{d4h}.  $D_4^h$ has eight
one-dimensional and two two-dimensional irreps.

In the continuum, the $T=0$ symmetry group is that of $O(3)$ rotations
of the slice, and breaks down for $T>0$ to the symmetry group of the
cylinder, ${\cal C}=O(2)\times Z_2$.  The lattice groups are subgroups of
these continuum groups.  The real irreps of $\cal C$ are easily obtained.
There are two one-dimensional irreps $0_+$ and $0_-$, which come from
the $J_z=0$ components of the even and odd spin representations of
$O(3)$. There is a countable infinity of two-dimensional real irreps
which span the spaces of $J_z=\pm M$ for any $J>0$. These also carry
irreps of the remaining $Z_2$ subgroup which corresponds to reflections,
$t\to-t$, in Euclidean time.

In the continuum limit the scalar irrep $A_1^+$ and a non-trivial irrep
$A_2^-$ of $D_4^h$ both collapse into the $0_+$ irrep of the cylinder
group; hence these screening masses must be degenerate in this limit.
The $0_-$ gets both the $A_1^-$ and $A_2^+$, and hence this pair
of screening masses must also become degenerate. The four remaining
one-dimensional irreps $B^\pm_{1,2}$ collapse into the $J_z=\pm2$ irrep
of $O(2)$, and the two-dimensional irreps $E^\pm$ become the $J_z=\pm1$
of $O(2)$. All these patterns of degeneracies have been verified in the
glue sector of the $SU(2)$ pure gauge theory \cite{saumen}.

These seemingly esoteric group theoretic constructions actually have
very simple physical applications. It has long been appreciated that
in screening phenomena, particles of different spin may mix, and that
different polarization components may have different dispersion relations
\cite{chin}. These correspond, respectively, to mixing of equal $J_z$
for different $J$ and the distinction between different $J_z$ for the
same $J$. While we discuss only physics for $T>0$ and zero chemical
potential here, it should be noted that the essential group theory lies
in the inequivalence of the spatial and temporal directions. This is
also true at finite chemical potential, and hence this group theory is
also relevant to such future lattice computations.

\subsection{Results}\label{sc.scr}

We have investigated the usual ``local'' staggered meson operators,
{\sl i.e.\/}, those in which the quark and the anti-quark both land on
the same lattice site (see Table \ref{tb.meson1}). These are familiar
from their $T=0$ characters.  The scalar ${\bf 1}^{++}$, S, and the
pseudo-scalar ${\bf 1}^{+-}$, PS, both belong to the $A_1^+$ of
$D_4^h$. The three components of the vector ${{\bf 3}''''}^{++}$, V,
and the axial-vector ${{\bf 3}''''}^{+-}$, AV, can be further reduced.
These give two different copies of $A_1^+$, and also a $B_1^+$.  All
the $A_1^+$ irreps have been measured extensively before.

We have also made the first ever $T>0$ investigation of a class of
``non-local'' staggered mesons, one in which the quark and the anti-quark
are separated by one link (see Table \ref{tb.meson1}). These are made
gauge invariant by including any product of links that starts on one
of these sites and ends on another.  The various operators lie in the
$A_2^-$ and $E^-$ irreps of $D_4^h$.

\begin{table}[htbp]
  \begin{center}\begin{tabular}{|c||c||c|c|c||c|c|c|}  \hline
  Irrep & $T/T_c$ & \multicolumn{3}{c||}{S/PS}
                  & \multicolumn{3}{c|}{V/AV} \\
  \hline
                 && $N_f=0$ & $N_f=2$ & $N_f=4$
                  & $N_f=0$ & $N_f=2$ & $N_f=4$ \\
  \hline
  $A_1^+$ & 1.5  & $4.21\pm0.01$ & $3.67\pm0.02$ & $3.51\pm0.01$
                 & $5.32\pm0.08$ & $5.44\pm0.08$   & $5.6\pm0.1$\\
          & 2.0  & $4.528\pm0.008$ & $4.08\pm0.01$ & $3.744\pm0.001$
                 & $5.64\pm0.02$ & $5.72\pm0.04$   & $5.6\pm0.1$\\
          & 3.0  & $4.84\pm0.04$ & $4.340\pm0.008$ & $4.29\pm0.01$
                 & $5.656\pm0.008$ & $5.72\pm0.03$   & $5.6\pm0.1$\\
  \hline
  $A_2^-$ & 1.5  & $4.6\pm1.1$ & $4.6\pm1.0$ & $\times$
                 & $4.4\pm1.1$ & $4.4\pm1.0$ & $\times$ \\
          & 2.0  & $4.4\pm0.8$ & $4.1\pm0.8$ & $4.1\pm1.2$ 
                 & $4.4\pm1.0$ & $3.6\pm1.2$ & $3.9\pm0.7$\\
          & 3.0  & $4.4\pm0.6$ & $4.8\pm0.8$ & $5.2\pm2.0$ 
                 & $5.0\pm0.6$ & --- & $5.6\pm2.8$ \\
  \hline
  \end{tabular}\end{center}
  \caption[dummy]{Screening masses in units of temperature, $M/T$, in various
     ``meson'' channels. A cross indicates that no data exists, and
     a dash that no stable measurement was possible. The $N_f=4$ results are
     from runs reported in \protect\cite{dimred}.}
\label{tb.masses}\end{table}

Chiral symmetry restoration gives the following relations between the 
staggered local meson correlators---
\beq
   C_{PS}(z) = (-1)^z C_S(z) \qquad{\rm and}\qquad
   C_{AV}(z) = (-1)^z C_V(z).
\label{corid}\eeq
We found that these relations are satisfied for almost all the correlators
to great accuracy at all $T$. The only small discrepancy is for the S/PS
correlators at $T=1.5T_c$, where the relation is violated at the 68\%
confidence level (but not at the 95\% confidence level). In order to rule
out the finite MD-time step errors as the source of this oddity, we ran a
second simulation at this coupling with half the time step. The results
remained unchanged. Since our error estimates are stable, we performed
another run with $N_z=24$. On this longer lattice the S and PS agreed
at the 68\% confidence level.

For staggered Fermions, correlators are usually fitted to a form such as
\beq
   C(z) = A_1 \cosh[M_1 (z-N_z/2)] + (-1)^z A_2 \cosh[M_2 (z-N_z/2)].
\label{fitf}\eeq
The correlator identities in (\ref{corid}) imply that the combinations
$C_V+C_{AV}$ and $C_V+(-1)^z C_{AV}$ project out the mass eigenstates in
the chiral symmetric phase. We have also made single mass fits to these
projections to check for stability of the fitted parameters. In addition
we have extracted screening masses by constructing local masses with
these projections. The local mass, $m(z)$, is defined by the solution of
\beq
   \frac{\cosh[m(z) (z-1-N_z/2)]}{\cosh[m(z)(z+1-N_z/2)]}
       = \frac{C(z-1)}{C(z+1)}\,,
\label{locm}\eeq
where the correlation functions are estimated by a jack-knife procedure.
In order to take care of covariance between various measurements, we
take the number of jack-knife bins to be much smaller than the number
of measurements.

In the $N_f=4$ theory the $B_1$ correlators are identically
zero \cite{b1}.  A statistical test of this hypothesis is made through the
usual
\beq
   \chi^2 = \sum_{ij} (y_i - h_i) \left(\sigma^{-1}\right)_{ij}
        (y_j - h_j),
\label{chioffd}\eeq
where $y_i$ are the measurements of the correlators at different $z$,
$\sigma$ is the covariance of these measurements, and the hypothesis being
tested is that $h_i=0$. With the data we have gathered in $N_f=2$ QCD,
we found that
\beq
   \chi^2 = \cases{15.8 & ($1.5T_c$)\cr
                   13.2 & ($2.0T_c$)\cr
                   24.1 & ($3.0T_c$)}     
\label{values}\eeq
Since $N_z=16$ in all these measurements, the number of degrees of
freedom is 15.  Thus, no alternative hypothesis can be supported by
our data at the 99\% confidence limit for $T\le2T_c$. For $T=3T_c$,
where there might be a signal, a one-mass fit by the first term of eq.\
(\ref{fitf}) gave a coefficient about $3\sigma$ away from zero, and a
screening mass $M/T=4.5\pm0.2$. In the quenched theory, on the other
hand, the situation was more like the $N_f=4$ theory. At all $T$ the
$B_1^+$ correlator was compatible with zero. The fits also gave similar
agreement with a vanishing correlator. More detailed studies of the $B_1$
correlator with high statistics at several temperatures above $2T_c$
might be of interest in the future.

The correlation functions in the remaining irreps are non-trivial.
The screening masses obtained are listed in Table \ref{tb.masses},
where we have also collected measurements made with quenched and $N_f=4$
dynamical staggered Fermions on $N_t=4$ lattices. Note that the two
sets of $A_1^+$ correlators, the S/PS and the V/AV, give different
screening masses. The V/AV screening mass, $M_V$, is consistent with
free field theory.  At all $T$, the $A_2^-$ screening mass is close to
the S/PS screening mass, $M_S$, but due to the large errors, is also
consistent with $M_V$.  The $E^-$ screening mass is harder to determine
because the correlator is very noisy; the central value is consistent
with free field theory, but has very large errors.

The two parity projected local V and AV correlators are identical.
This implies that not only the screening masses, but also the mixture of
states excited by the operator are identical. In other words, the $a_1$
and $b_1$ screening states mix for $T>T_c$. Similar equalities were
found in the parity projections of the $A_2^-$ (and $E^-$) correlators
coming from the V and AV non-local meson operators, showing that the
$\rho$ and $\omega$ screening states also mix.  The non-local S and PS
correlators similarly bear evidence for the mixing of $\pi$ and $a_0$. In
addition, the equality of the $A_2^-$ screening mass determined from
all four $A_2^-$ correlators at our disposal also argues for a mixing
of $\pi$ and the $J_z=0$ component of the $\rho/\omega$. Thus both the
phenomena expected in screening correlators are seen--- mixing of states
of different $J$ and different components of the same $J$ having different
dispersion relations.

\section{Quark number susceptibilities}

\subsection{Definitions}

We define the partition function
\beq
   Z = \int{\cal D}U {\rm e}^{-S_g}
            \det M(m_u,\mu_u)\det M(m_d,\mu_d)\det M(m_s,\mu_s),
\label{part}\eeq
where $S_g$ is the gluon action of interest and $M$ denotes
appropriate lattice Dirac operators.
The chemical
potentials for each flavor can be combined into the singlet, triplet
and octet $SU(3)$ chemical potentials---
\beq
   \mu_0=\mu_u+\mu_d+\mu_s,\qquad
   \mu_3=\mu_u-\mu_d,\qquad{\rm and}\qquad
   \mu_8=\mu_u+\mu_d-2\mu_s.
\label{su3}\eeq
Our notational convention is that indices such as $f$ and $f'$ run over
the flavors $u$, $d$ and $s$, and indices such as $i$ and $j$ run over
the $SU(3)$ diagonal generators 0, 3 and 8.

The quark number densities are
\beq
   n_f \equiv \left(\frac{T}{V_3}\right)\frac{\partial\ln Z}{\partial\mu_f}
       = \left(\frac{T}{V_3}\right)
	     \left\langle{\rm tr} M_f^{-1} M_f'\right\rangle,
\label{number}\eeq
where $M_f'=\partial M_f/\partial\mu_f$, $V_3 = N_x^3a^3$, and $T
= 1/N_ta$.  Conversion to the other basis simply follows using the
definitions in eq.\ (\ref{su3}) and the chain rule of differentiation.
Note that $n_0=(n_u+n_d+n_s)/3$ and $n_3=(n_u-n_d)/2$ are the baryon number
and isospin densities, respectively. These are densities of (conserved)
charges and not of quark number.  The quark number susceptibilities are
the second derivative of the free energy with respect to the chemical
potentials
\beq
   \chi_{ff'} \equiv \frac{\partial n_f}{\partial \mu_{f'}}
             = \left(\frac{T}{V_3}\right)
      \left[\frac1Z\frac{\partial^2Z}{\partial\mu_f\partial\mu_{f'}}
          -\frac1Z\frac{\partial Z}{\partial\mu_f}\,
           \frac1Z\frac{\partial Z}{\partial\mu_{f'}}\right].
\label{incomplete}\eeq
To lighten the notation, we shall put only one subscript
on the diagonal parts of $\chi$.

We are interested in evaluating the susceptibilities for $m_u=m_d<m_s$
at the point $\mu_f=0$ for all $f$, yielding much simplification.
For example, each $n_f$ vanishes, a fact that we utilize as a check on
our numerical evaluation.  Moreover, the product of the single derivative
terms in eq.\ (\ref{incomplete}) vanishes, since each is proportional
to a number density.  Finally, since the masses are degenerate,
$M(m_u,0)=M(m_d,0)$ for each configuration of gauge links under study.

Flavor off-diagonal susceptibilities such as
\beq
   \chi_{ud} = \left(\frac{T}{V_3}\right)
         \left\langle{\rm tr} M_u^{-1} M_u'\;
                     {\rm tr} M_d^{-1} M_d'\right\rangle
\label{offd}\eeq
are given entirely in terms of the expectation values of disconnected
loops. Such quantities are numerically hard to compute. We discuss their
evaluation later. Since $M_u=M_d$, we obtain
$\chi_{us}=\chi_{ds}$ with each defined by an obvious generalization
of the formula above. Of the flavor diagonal susceptibilities we shall use
\beq
   \chi_s = \left(\frac{T}{V_3}\right) \left[
         \left\langle\left({\rm tr} M_s^{-1} M_s'\right)^2
                     \right\rangle +
         \left\langle{\rm tr}
             \left(M_s^{-1} M_s'' - M_s^{-1}M_s'M_s^{-1}M_s'\right)
                  \right\rangle\right].
\label{chis}\eeq
$\chi_u=\chi_d$ are given by an obvious generalization of this formula.
Numerically, the simplest quantity to evaluate is the diagonal iso-vector 
susceptibility
\beq
   \chi_3 = \frac12 \left(\frac{T}{V_3}\right) 
         \left\langle{\rm tr} \left(M_u^{-1} M_u'' 
                 - M_u^{-1}M_u'M_u^{-1}M_u'\right)\right\rangle.
\label{chi3}\eeq

Two more susceptibilities are of interest. These are the baryon number and 
electric charge susceptibilities, 
\beq
   \chi_0 = \frac19\left(4\chi_3+\chi_s+4\chi_{ud}+4\chi_{us}\right)
       \qquad{\rm and}\qquad
   \chi_q = \frac19\left(10\chi_3+\chi_s+\chi_{ud}-2\chi_{us}\right).
\label{chi0q}\eeq
Note that $\chi_0$ is the baryon number
susceptibility for three flavors of quarks. As a result, this expression
differs from the iso-singlet quark number susceptibility for two flavors,
defined in \cite{qsdef}, both in overall normalization and by terms
containing strangeness.

Note that quark masses appear in two places: first in the determinant
in (\ref{part}) which defines the weights for the averaging, and second
in the trace of the Dirac operators which define the susceptibilities.
The first is the dynamical sea quark mass $m$, and the second is the
valence quark mass $m_v$.  In principle these can be different. We
work with light sea quark masses for the u and d flavors, as described
earlier, and use the quenched approximation for the strange quark,
{\sl i.e.\/}, set $\det M_s = 1$.  We have earlier reported measurements
of these susceptibilities when all the flavors are quenched \cite{qsus}.
For staggered quarks each trace in eqs.\ (\ref{offd}--\ref{chi3}) comes
with a factor of $1/4$ to compensate for flavor doubling. We differ from
the conventions of \cite{qsdef} by this overall factor.

These susceptibilities are easy to compute in free field theory {\sl i.e.\/},
for an ideal Fermi gas. As expected, $n_f=0$ for $\mu_f=0$.  All the
off-diagonal susceptibilities are also zero. For $\chi_3$ we obtain
\beq
   \chi_{FFT} = \frac1{4N_tN_x^3}\sum_p
     \frac{\sin^2p_0\cos^2p_0}{\left[m^2+\sum_\nu \sin^2p_\nu\right]^2},
\label{sus}\eeq
where the spectrum of momenta is $p_0=(2\pi/N_t)(n_0+1/2)$,
for $n_0=0,\cdots,N_t-1$, and $p_\nu=(2\pi/N_x) n_\nu$, where
$n_\nu=0,\cdots,N_x-1$ for $\nu\ne0$. 
$\chi_s$ is also given by the same expression for the appropriate quark
mass. $\chi_0^{FFT}$ and $\chi_q^{FFT}$ can then be obtained using eq.\
(\ref{chi0q}).

\subsection{Optimizing the measurements}

The traces required in measurements of the quark number susceptibility
were estimated using a well-known stochastic method. Such stochastic
estimators for the traces of $N\times N$ matrices can be constructed
from ensembles of $N$-dimensional complex vectors, $R$, of which each
component, $z_\alpha$, is drawn from a Gaussian distribution of unit
variance. From the identity
\beq
   \frac12\int z^*_\alpha z_\beta \prod_{\mu=1}^N {\rm e}^{-|z_\mu|^2/2}
      \left(\frac{d^2z_\alpha}{2\pi}\right) = \delta_{\alpha\beta},
\label{id0}\eeq
(here $d^2z$ means $rdrd\theta$ and the complex number $z=r\exp i\theta$)
we construct the estimator by multiplying both sides by a matrix
element $A_{\alpha\beta}$ and summing over the indices. The right
hand side gives ${\rm tr}A$. The integral on the left hand side
has an obvious Monte Carlo estimator, giving the relation
\beq
   {\rm tr} A = \frac1{2N_v}\sum_{i=1}^{N_v} R_i^\dag A R_i
      = \overline{R^\dag AR},
\label{cstoch}\eeq
where $N_v$ is the number of vectors used in making the estimate. Note that
eq.\ (\ref{id0}) implies that the factor 2 above is part of the weight in
the average over the ensemble of random complex vectors.

\begin{table}[hbtp]
\begin{center}\begin{tabular}{|c||c|c|c|c||c|c|c|}\hline
 $m_v/T_c$ & \multicolumn{4}{c||}{$4\times8^3$}
           & \multicolumn{3}{c|}{$4\times16^3$} \\
\hline
 & $N_v$  & $\chi_3/T^2$ &$10^4\chi_{ud}/T^2$ & $\chi_\pi/T^2$
          & $\chi_3/T^2$ &$10^4\chi_{ud}/T^2$ & $\chi_\pi/T^2$ \\
\hline
  0.1   & 20 & 0.96 (4)  & $-2$ (2)   &104 (6) & & &\\
        & 40 & 0.96 (3)  & $-2.1$ (6) & 
             & 0.98 (3)  &  0.02  (2) &106 (6) \\
        &100 & 0.96 (2)  &  0.06  (4) & & & &\\
  0.3   &100 & 0.91 (1)  &  0.06  (4) & 96 (4)
             & 0.91 (2)  &  0.02  (2) & 96 (3) \\
  0.5   &    & 0.84 (1)  &  0.05  (3) & 88 (3)
             & 0.83 (1)  &  0.02  (1) & 87 (2) \\
  0.75  &    & 0.748 (9) &  0.05  (3) & 78 (2)
             & 0.726 (8) &  0.02  (1) & 78 (2) \\
  1.0   &    & 0.660 (7) &  0.04  (3) & 70 (2)
             & 0.639 (6) &  0.014 (8) & 70 (2) \\
\hline
\end{tabular}\end{center}
\caption{Susceptibilities for $T=1.5T_c$ and $m/T_c=0.1$. On the larger
   lattice we used $N_v=40$. $\chi_\pi$ is measured with a single point
   source by integrating the local PS correlator.}
\label{tb.sus2}\end{table}

The stochastic estimator for $({\rm tr}A)^2$ is a small modification of
that given in \cite{qsdef}. We draw $L$ sets of independent random Gaussian
vectors as before and construct the estimator
\beq
   ({\rm tr} A)^2 = \frac2{L(L-1)}\sum_{i>j=1}^L
            \left(\overline{R_i^\dag A R_i}\right)\left(\overline{R_j^\dag A R_j}\right).
\label{id3}\eeq
If we draw $N_v$ vectors in the evaluation of the single trace, then
it is simplest to divide these into $L$ sets of $N_v/L$ vectors for the
estimate of each ensemble average in the sum above \footnote{We thank
K.\ Kanaya for a discussion on this estimator.}.

In \cite{qsdef} the sum above was taken over all pairs, and the diagonal
term was removed by subtracting a term equal to a stochastic estimator
for ${\rm tr}(A^2)$. In \cite{qsus} where that method was used, we showed
that reasonably large values, $N_v\approx100$ were necessary to avoid a
systematic bias in $\chi_0-\chi_3$.  However, the estimate is facilitated
by excluding the diagonal term from the sum above.  For example, we found
that the error in ${\rm tr} A$ scales as $1/\sqrt{N_v}$, whereas that in
$({\rm tr} A)^2$ scales as $1/N_v$, for fixed $L$. The dependence on $L$
was marginal, as long $N_v/L>5$.

\begin{table}[htbp]
\begin{center}\begin{tabular}{|c||c|c|c|c|c||c|c|c|c|c|}\hline
$m_v/T_c$ & $T/T_c$ & $\chi_{\scriptscriptstyle FFT}/T^2$ & $\chi_3/T^2$ &$10^6\chi_{ud}/T^2$ & $\chi_\pi/T^2$ &
            $T/T_c$ & $\chi_{\scriptscriptstyle FFT}/T^2$ & $\chi_3/T^2$ &$10^6\chi_{ud}/T^2$ & $\chi_\pi/T^2$ \\
\hline
0.1 &1.25&1.1274& 0.90 (3)  & 0   (1) & 149 (10) &
     2.00&1.1276& 0.973 (7) & 0   (1) & 84 (2) \\
0.3 &    &1.1250& 0.80 (1)  & 0.4 (8) & 119 (6) &
         &1.1227& 0.951 (7) & 0   (1) & 82 (2) \\
0.5 &    &1.1203& 0.704 (8) & 0.2 (7) &  99 (4) &
         &1.1248& 0.888 (9) & 1   (1) & 79 (2) \\
0.75&    &1.1112& 0.598 (6) & 0.0 (6) &  83 (2) &
         &1.1212& 0.849 (6) & 0.3 (8) & 75 (1) \\
1.0 &    &1.0988& 0.511 (6) & 0.1 (5) &  72 (2) &
         &1.1162& 0.783 (6) & 0.4 (7) & 70 (1) \\ \hline
0.1 &1.50&1.1275& 0.96 (1)  & 4   (3) &  99 (4) &
     3.00&1.1277& 0.995 (4) & 0.7 (7) & 75 (1) \\
0.3 &    &1.1259& 0.903 (7) & 3   (2) &  93 (3) &
         &1.1273& 0.988 (4) & 0.7 (7) & 75 (1) \\
0.5 &    &1.1226& 0.829 (6) & 3   (2) &  85 (3) &
         &1.1264& 0.973 (4) & 0.7 (7) & 74 (1) \\
0.75&    &1.1162& 0.733 (4) & 2   (1) &  76 (2) &
         &1.1248& 0.946 (4) & 0.6 (6) & 72 (1) \\
1.0 &    &1.1074& 0.646 (4) & 1.8 (9) &  69 (1) &
         &1.1226& 0.912 (4) & 0.6 (6) & 70 (1) \\
\hline
\end{tabular}\end{center}
\caption{Susceptibilities measured on $4\times12^3$ lattices for $N_f=2$
   with $m/T_c=0.1$ and $N_v=100$. $\chi_\pi$ is measured with a single
   point source by integrating the local PS correlator.}
\label{tb.sus1}\end{table}

\begin{table}[hbtp]
\begin{center}\begin{tabular}{|c|c|c|c|c|c|c|}\hline
 $m_v/T_c$ & $\chi_{\scriptscriptstyle FFT}/T^2$ & $\chi_3/T^2$ &$10^6\chi_{ud}/T^2$ & $\chi_\pi/T^2$ \\
\hline
     0.075 &1.1276& 0.97 (1)  & 2   (2) & 106 (5) \\
     0.1   &1.1275& 0.97 (1)  & 1   (2) & 106 (4) \\
     0.3   &1.1259& 0.906 (8) & 1   (1) &  98 (4) \\
     0.5   &1.1226& 0.828 (6) & 0   (1) &  90 (3) \\
     0.75  &1.1162& 0.729 (5) & 0.4 (8) &  80 (2) \\
     1.0   &1.1074& 0.641 (4) & 0.2 (6) &  72 (2) \\
\hline
\end{tabular}\end{center}
\caption{Susceptibilities measured at $T=1.5T_c$ on $4\times12^3$ lattices
   for $N_f=2$ with $m/T_c=0.075$ and $N_v=80$.}
\label{tb.sus1p}\end{table}

\begin{table}[hbtp]
\begin{tabular}{|c|c|c|c|c|}
\hline
$m_v/T_c$ &  $\chi_3/T_c^2$ & $10^6\chi_{ud}/T_c^2$ & $\chi_{\pi}/T_c^2 $\\
\hline
 0.10 & 0.18 (7)  & 0   (1) & 462 (11)  \\
 0.30 & 0.024 (9) & 0.1 (5) & 182 (3)  \\
 0.50 & 0.007 (4) & 0.1 (3) & 122 (2) \\ 
 0.75 & 0.002 (2) & 0.1 (2) &  89 (1) \\ 
 1.00 & 0.000 (1) & 0.0 (1) & 71.4 (8) \\    
\hline
\end{tabular}
\caption{Susceptibilities at $T=0$ in units of $T_c$, measured on $12^4$
   lattices for $N_f=2$ with $m/T_c$=0.1 and $N_v=80$}.
\label{tb.zero}\end{table}  

Since the simulation time increases at least linearly in volume, we
need to make a choice of volume which reduces computation time without
introducing large artifacts into the measurements. We have made a
detailed study of the volume dependence of susceptibility measurements
at $T=1.5T_c$.  As the data in Table \ref{tb.sus2} clearly show, there is
no significant volume dependence in the values of $\chi_3$ and $\chi_{ud}$
in going from $4\times8^3$ to $4\times16^3$ lattices.

In view of this, we have chosen to make the remaining measurements
on $4\times12^3$ lattices. Close to $T_c$ we would expect that the
measurements are strongly volume dependent. Based on our estimates
of the screening masses we find that even at our smallest temperature
of $1.25T_c$ the lattice is more than 10 times larger than the longest
correlation length. This assures us that finite lattice size effects are
negligible at the lower end of our temperature range. On the other end of
the scale, by the simple expedient of not going beyond $3T_c$, we avoid
the finite volume effects such as the onset of spatial deconfinement.

Finally, one must optimize the conjugate gradient stopping criterion,
in which the norm of the residual vector is required to be smaller
than $\epsilon\sqrt V$. We were pleasantly surprised to find that, in
thermalized test configurations, results for $\chi_3$ and $\chi_{ud}$
changed by less than 1 part in $10^4$ on changing $\epsilon$ in the
range from $0.01$ to $10^{-6}$.  Increasing $\epsilon$ by one order
of magnitude meant a decrease of CPU time by 20--25\%. These numbers
were reproduced for test configurations on three lattice volumes, and a
variety of $T$. In actual measurements we chose $\epsilon$ to be $10^{-5}$
for $T\ge1.25T_c$ and $10^{-3}$ otherwise.

\subsection{Results}

\begin{figure}[hbt]
   \includegraphics{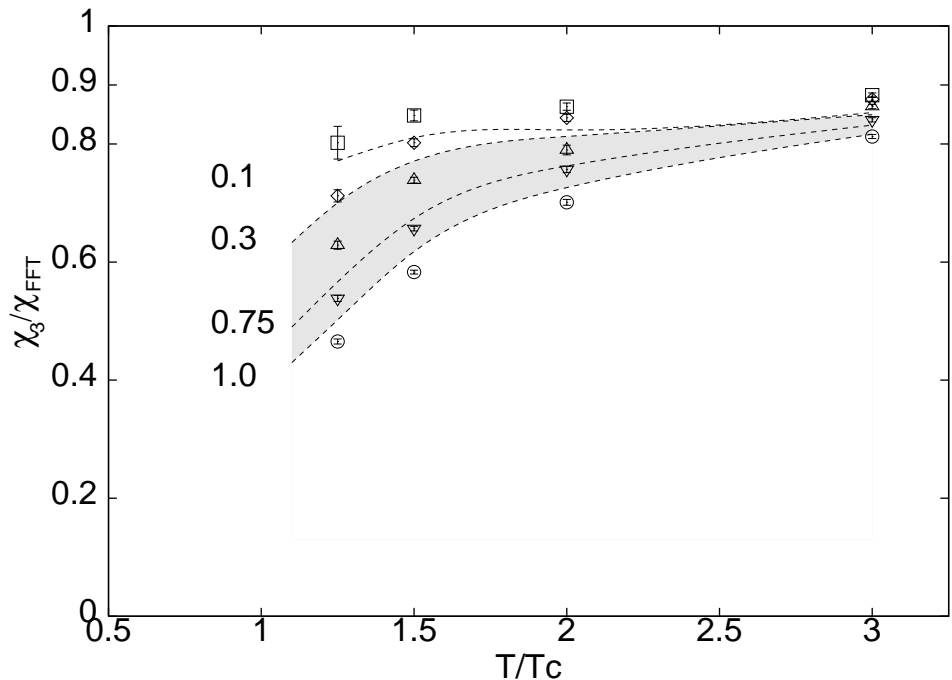}
   \caption{$\chi_3/\chi_{FFT}$ as a function of $T/T_c$ for $m_v/T_c=0.1$
      (boxes), 0.3 (diamonds), 0.5 (up triangles), 0.75 (down triangles) and 1
      (circles); all for $m/T_c=0.1$. For comparison the curves show the
      measurements in quenched QCD. All the data shown here is taken on
      $4\times12^3$ lattices. The shaded area denotes the mass range for
      strange quarks.}
\label{fg.chi3}\end{figure}

Our main results on the $T$ and $m_v$ dependence of $\chi_3$ and
$\chi_{ud}$ are collected in Table \ref{tb.sus1} for two flavors of
dynamical Fermions with $m/T_c=0.1$.  The results of a simulation
at $T=1.5T_c$ with a smaller $m/T_c=0.075$ are collected in Table
\ref{tb.sus1p}.  Comparing the two, we see no statistically significant
change in $\chi_3$ over this range of masses.  While $\chi_3$ is expected
to be strongly dependent on $m$ for $T\simeq T_c$ in the chiral limit,
our results indicate that the dependence of $\chi$ on the sea quark
mass away from the critical region is small. This suggests that the
values of $\chi_3$ or $\chi_{ud}$ in the chiral limit for $T\ge1.25T_c$
are within errors those in Table \ref{tb.sus1}. Table \ref{tb.zero}
shows that all the susceptibilities indeed vanish at $T=0$ as expected.

A comparison of the dynamical and quenched theories is displayed in
Figure \ref{fg.chi3} in terms of the ratio $\chi_3/\chi_{FFT}$. While the
general trend of the data are similar in the two cases, some quantitative
differences are visible. The most important of these is in the asymptotic
value of $\chi/\chi_{FFT}$.  For small $m_v$ the ratio is 0.85 at $T/T_c$
= 3 in the quenched theory, whereas it is 0.88 for $N_f=2$. For larger
$m_v$, $\chi_3$ is almost unchanged on unquenching the Fermions. $\chi_s$
changes by less than 3\% when the light fermions are unquenched, leading
us to believe that the unquenching of the strange quark will not affect
this quantity significantly.

These observations are not explained in continuum perturbation theory at
high temperature, which yields
\beq
   \chi/\chi_{FFT} = 1 - 2 \left(\frac\alphas\pi\right)
       + 8 \sqrt{1+\frac{N_f}6} \left(\frac\alphas\pi\right)^{3/2},
\label{pert}\eeq
when the plasmon term is resummed \cite{pert}. Here $\alphas$ is the
strong coupling at a scale appropriate to the temperature $T$.  It is easy
to see that this ratio is never less than $0.981$ for $N_f=0$ ($0.986$
for $N_f=2$).  This minimum is reached when $\alphas=\pi/6(6+N_f)$. A
recent analysis showed that $T_c/\Lambda_{\overline{MS}}=1.15\pm0.05$
for $N_f=0$ and $0.49\pm0.02$ for $N_f=2$ \cite{tc2}. Taking the scale
for the strong coupling to be $2\pi T$, this means that for $N_f=2$
the minimum occurs at $T\simeq3300\Lambda_{\overline{MS}}$ (and
$T\simeq110\Lambda_{\overline {MS}}$ for $N_f=0$). Thus in both cases,
the weak coupling estimates decrease as a function of temperature in the
range studied here, in contrast to our lattice results, which increase.
It would be interesting to check whether a full ${\cal O}(\alphas^{5/2})$
computation for $\chi$ approaches the lattice results, and whether any
resummations of the perturbation theory do better \cite{resum,toni}.

\begin{figure}[!htb]
   \includegraphics{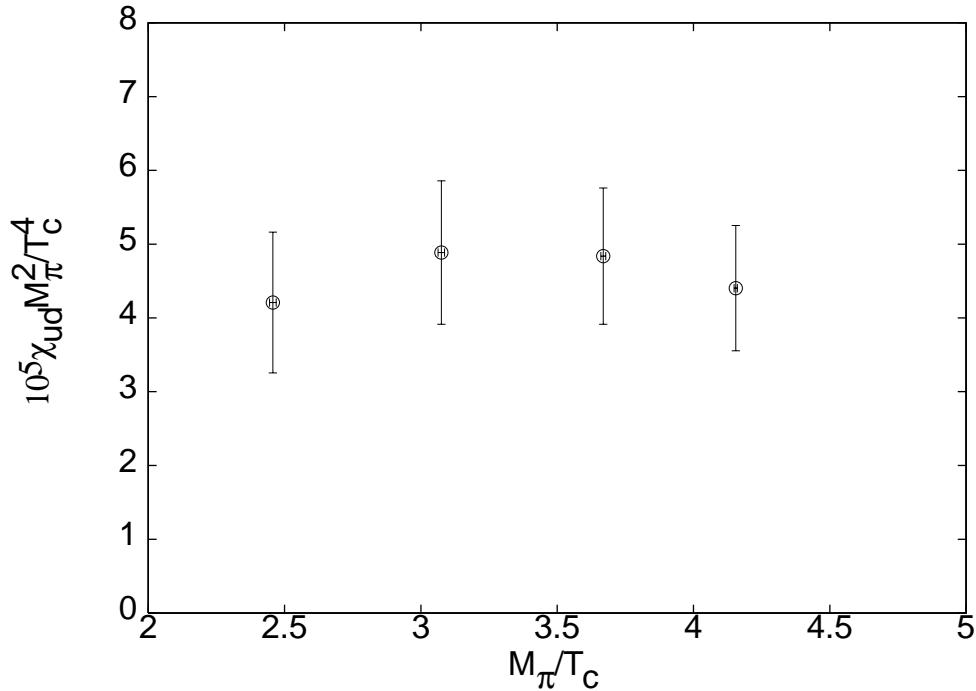}
   \caption{$\chi_{ud}$ scales as $1/M_\pi^2$ for $T=0.75T_c$ in
      quenched QCD. The measurements of $\chi_{ud}$ and $M_\pi$ were
      both performed on the same configurations on $4\times12^3$
      lattices \cite{qsus}.}
\label{fg.offd}\end{figure}

The off-diagonal susceptibility, $\chi_{ud}$, vanishes for $T>T_c$
and also at $T=0$. As shown in Table \ref{tb.sus1}, the errors on
$\chi_{ud}/T^2$ for $T>T_c$ are of the order of $10^{-6}$, and hence
this is the precision within which this quantity can be said to vanish
for all $m_v$. Thus, $\chi_{ud}$ and other non-diagonal parts of the
flavor space susceptibilities can be totally neglected in constructing
other susceptibilities. It is interesting to note that $\chi_{ud}$
is zero in an ideal gas, and perturbative contributions start at
order $\alphas^2$. Taking the scale to be $2\pi T$, as before, and
the same values of $T_c/\Lambda_{\overline MS}$, we might then expect
$\chi_{ud}$ to be non-zero at the level of $0.04$--$0.1$ (for $N_f=2$)
in the temperature range we studied.  The substantially smaller, indeed
vanishing, value thus accentuates the puzzle about perturbation theory
\footnote{In \protect\cite{toni}, it is also shown that $\chi_{ud}$
starts at order $\alphas^3$ with a small coefficient, bringing the
resummed prediction closer to the data.}.

The only place where $\chi_{ud}$ is definitely non-zero is for $0<T<T_c$.
A re-analysis of our data in the quenched theory at $T=0.75T_c$
\cite{qsus} showed that $\chi_{ud}$ differs from zero by over four
standard deviations.  As displayed in Figure \ref{fg.offd}, $\chi_{ud}$
scales inversely with the square of the pion screening mass, $M_\pi$, at
this temperature. Since it is known that the hadron masses are independent
of temperature from $T=0$ to some point rather close to $T_c$ \cite{lowt},
the only possible temperature dependence of $\chi_{ud}$, for such $T$,
would be in the value of the constant $\chi_{ud} M_\pi^2$. Note that the
spatial lattice size is such that $M_\pi L\ge10$, and finite size effects
are negligible. In order to ensure this on $4\times12^3$ lattices, we
are constrained to use reasonably heavy quarks. The pion screening mass
is then rather large. In future we plan to study $\chi_{ud}$ for $T<T_c$
in more detail by pushing towards lower pion mass with larger lattices,
covering a bigger range of $T$ and using dynamical quarks.

\subsection{Relation to screening masses}\label{sc.chi}

\begin{figure}[!htb]
   \includegraphics{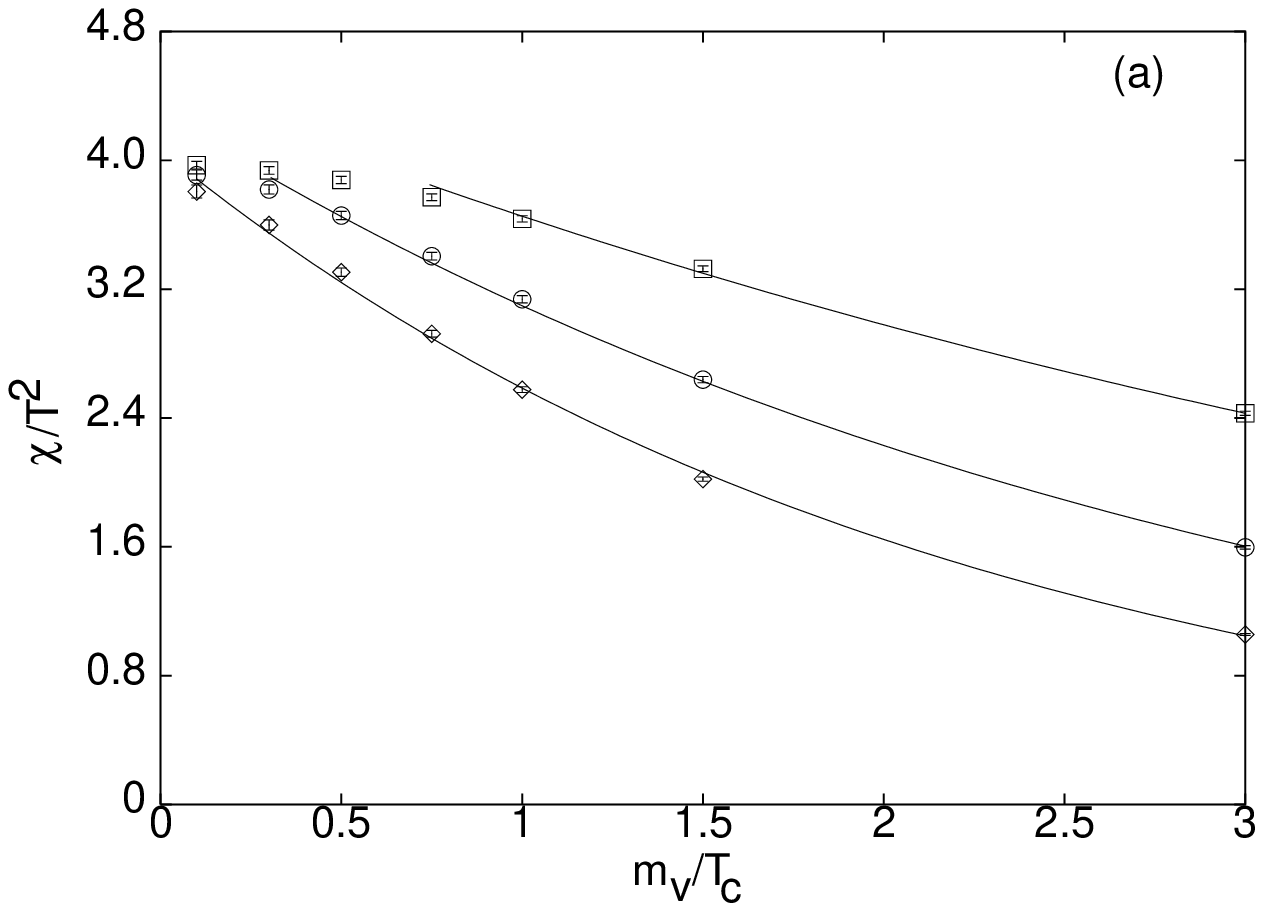}
   \includegraphics{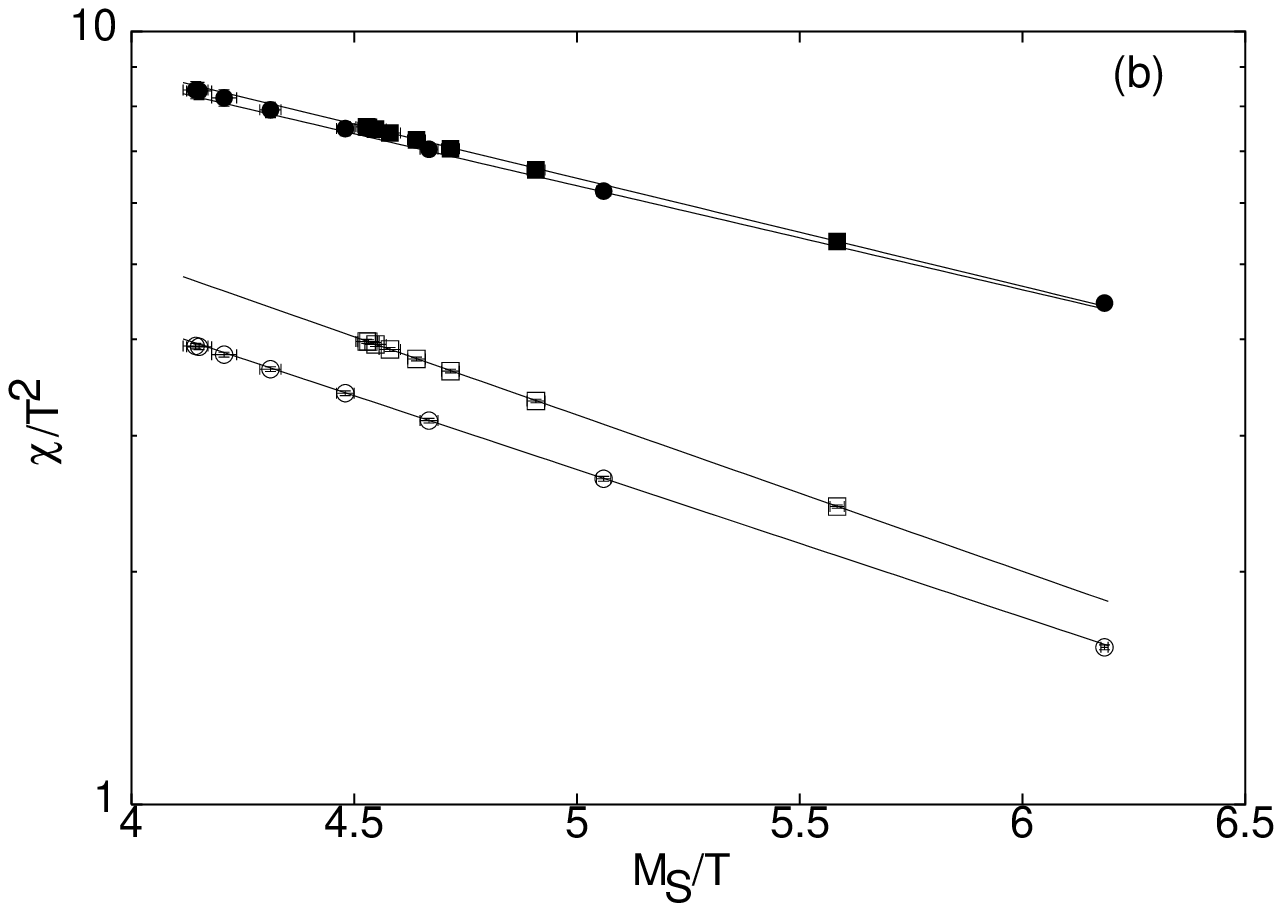}
   \caption{Panel (a) shows $4\chi_3/T_c^2$ as a function of $m_v/T_c$ (for
      sea quark mass of $0.1T_c$) at $T=1.5T_c$ (diamonds), $2T_c$ (circles)
      and $3T_c$ (boxes). Panel (b) shows $4\chi_3/T^2$ (open symbols) and
      $\chi_\pi/10T^2$ (filled symbols) as a function of $M_S/T$ at $2T_c$
      (circles) and $3T_c$ (boxes). The lines are exponential fits.}
\label{fg.shape}\end{figure}

The physics of a plasma lies in screening. However, it is not always
clear how screening appears in different guises. Interestingly it can be
shown that quark number susceptibility is directly related to screening,
and therefore the non-perturbative physics in the two are likely to be
the same.  It is possible to present a deductive argument from first
principles and use the data to illustrate it. Instead, we first show how
the data prefers a relation between $\chi$ and the S/PS screening mass
$M_S$, and then give the group theoretical argument that this is not a
numerical coincidence.

The variation of $\chi_3$ with $m_v$ at fixed $T$ is shown in Figure
\ref{fg.shape}. For large enough $m_v$, there is an exponential fall
of $\chi_3$, which can be qualitatively understood as the effect of
Boltzmann factors. It is less obvious that there should be a threshold
$m^*(T)$ below which $\chi_3$ is almost independent of $m_v$. The data
indicates that $m^*(T)\simeq T$.  Such a threshold cannot be derived in
a weak coupling expansion.  Furthermore, this threshold of constancy
is exactly what prevents $\chi_3/\chi_{FFT}$ from reaching unity as
$m_v\to0$, since Figure \ref{fg.shape} is just another representation
of the data in Figure \ref{fg.chi3}.

The only other observed non-perturbative effect in the quark sector
of high-temperature QCD is in the screening mass in the S/PS channel,
$M_S$. That these two effects are related is shown in the second panel of
Figure \ref{fg.shape}, where we plot $\chi_3$ against $M_S/T$. A simple
exponential relation between them is seen. The threshold we saw before
disappears into the relation between $m_v$ and $M_S$. The remaining
question is why the two should be related at all.

The answer is simple, and relies on the group theoretical classification
of screening correlators and masses according to the symmetries of the
$T>0$ transfer matrix in the spatial direction \cite{d4h}.  First,
from eq.\ (\ref{chi3}) it can be seen that $\chi_3$ is the thermal
expectation value of a product of two quark propagators sandwiching
$\gamma_0 U$ and summed over all distances. As a result, this quantity
can be related to a susceptibility constructed from a one-link separated
meson screening correlator. In the $T=0$ notation, this turns out to be
a component of the one-link V ${\bf 3}^{-+}$. For $T>0$ this component
reduces to the irrep $A_2^-$ of the appropriate symmetry group $D_4^h$
(see Table \ref{tb.meson1}). Other correlators which lie in the same
irrep are the one-link separated S and PS ${{\bf 3}''}^{-\pm}$ and the a
specific component of the one-link AV ${\bf 3}^{--}$.  In the continuum
limit the spectrum of $A_2^-$ screening masses is degenerate with that
of the $A_1^+$, since they come from the same irrep of the continuum
$O(2)$ symmetry \cite{saumen}. The smallest $A_1^+$ screening mass comes
from the S/PS local propagators which are used to extract $M_S$. This,
is the reason for the close relation between $\chi_3$ and $M_S$ shown
in Figure \ref{fg.shape}.

Further numerical evidence in favor of this group theoretical argument is
the similarity in the relation between $\chi_\pi(M_S)$ and $\chi_3(M_S)$,
as shown in Figure \ref{fg.shape}. The near equality of the slopes lends
support to our earlier observation that the $A_2^-$ screening mass
seems to be equal to $M_S$. We expect that for smaller lattice spacing
or on using a Fermion action which restores the full flavor symmetry,
the two screening masses should become equal. In this limit, the lines
would be exactly parallel and the difference in values of $\chi_\pi$
and $\chi_3$ at the same $M_S$ would only reflect different operator
overlaps with the same state. This, in fact, is a prediction which should
be tested in future.

One further piece of information fits neatly into this group theoretic
framework. $\chi_{ud}$ has the same symmetry as $\chi_3$ and therefore
it is also related to the S/PS correlators at all non-zero temperatures.
Since it vanishes for $T>T_c$, the relation is trivial there. However,
for non-zero $T<T_c$, we have observed that $\chi_{ud}=Z/M_\pi^2$. This
relation would be as mysterious as that between $\chi_3$ and $M_S$
in the absence of the argument given above.

It is interesting to note that the preceding arguments rest entirely
on the group theory of screening correlators and masses, {\sl i.e.\/},
on the spatial direction transfer matrix of the finite temperature
system. In terms of the Euclidean time direction transfer matrix,
the group theory is the same as at $T=0$ and there is no reason for
the one-link separated vector temporal correlator to be related to the
S/PS. Thus, the physics of quark number susceptibility is related to
screening correlators and not to temporal correlators.

\section{Discussion and Summary}

\subsection{Event-to-event fluctuations}

\begin{table}[!bhtp]
\begin{center}\begin{tabular}{|c|c|c|c|c|}\hline
 $T/T_c$ & $\chi_3/\chi_3^{FFT}$ & $\chi_s/\chi_s^{FFT}$ & $\chi_0/\chi_0^{FFT}$ & $\chi_q/\chi_q^{FFT}$ \\
\hline
 1.25 & 0.80  (3) & 0.538 (5) & 0.71  (2) & 0.76  (2) \\
 1.50 & 0.848 (9) & 0.657 (4) & 0.785 (6) & 0.817 (8) \\
 2.00 & 0.863 (6) & 0.757 (6) & 0.828 (5) & 0.846 (5) \\
 3.00 & 0.883 (4) & 0.841 (4) & 0.869 (3) & 0.877 (3) \\
\hline
\end{tabular}\end{center}
\caption{Susceptibilities (in units of their values for an ideal gas)
   obtained for degenerate dynamical u and d quarks of mass 17 MeV and
   quenched strange quarks with $m=125$ MeV. Susceptibilities below
   $T_c$ are discussed in the text.}
\label{tb.charge}\end{table}

Since $\chi_q$ is related to charge fluctuations, it can be observed
experimentally in heavy-ion collisions \cite{koch}.  The baryon
susceptibility, $\chi_0$, and the strange quark susceptibility, $\chi_s$,
can also be used to measure fluctuations of the corresponding quantities
\cite{chi0s}. Also, $\chi_s$ is related by a fluctuation-dissipation
theorem to the total amount of strangeness produced in equilibrium,
and hence can be used to probe departures from equilibrium in heavy-ion
collisions.  Given the relevance of these quantities to the experimental
search for the quark-gluon plasma, we have collected our results in
an easily usable form in Table \ref{tb.charge}.

The behavior of both $\chi_q$ and $\chi_0$ above $T_c$ are essentially
controlled by $\chi_3$. It is clear from eq.\ (\ref{chi0q}), that for
$T\gg T_c$, where $\chi_3\simeq\chi_s\gg\chi_{ud}\simeq\chi_{us}$, we must
have $\chi_q/\chi_0\approx2$. Closer to $T_c$, but still in the hot phase,
if $\chi_s$ is also neglected, we would have $\chi_q/\chi_0\approx5/2$. In
fact, we find that this ratio decreases from $2.13\pm0.08$ at $1.25T_c$
to $2.02\pm0.01$ at $3T_c$. This shows that $\chi_s$ is significant at all
temperatures in the plasma, but certainly becomes closer to $\chi_3$ with
increasing $T$ (Figure \ref{fg.chi3}). In any case, charge fluctuations
are twice as large as baryon number fluctuations. Interestingly, we
find that $\chi_q/\chi_s$ also falls from $0.95\pm0.03$ at $T=1.25T_c$
to $0.696\pm0.004$ at $T=3T_c$ (it is 2/3 in the $T \to \infty$ limit);
therefore strangeness fluctuations become more significant than charge
fluctuations with increasing $T$.

In the low-temperature phase, $0<T<T_c$, fluctuations are dominated
by the flavor off-diagonal contribution. We have displayed evidence
(Figure \ref{fg.offd}) that $\chi_{ud}\simeq1/m_\pi^2$, {\sl i.e.\/},
fluctuations are due to the lightest particle of given flavor. Although
the value of $\chi_{ud}$ is extremely small, the measure of fluctuation
is the ratio of $\chi$ and the entropy density \cite{koch}.  While there
are no reliable measurements of the entropy for $T<T_c$, it is known to
be small. It is presently an open question whether the ratio is smaller
for $T<T_c$ or on the other side of the QCD phase transition.

Assuming that the inverse relation between $\chi_{ud}$ and the pion
mass generalizes to other flavor off-diagonal susceptibilities, eq.\
(\ref{chi0q}) can be used to predict that $\chi_q/\chi_0\approx1/4$
for $T<T_c$.  Corrections to this number are then given in terms of
$(m_\pi/m_K)^2$ and $(m_\pi/m_\eta)^2$. These might further lower the
ratio by 15--20\%.  As a result, in the low-temperature phase we would
obtain the hierarchy of fluctuations--- $\chi_0>\chi_q>\chi_s$, in total
contrast to the inverted hierarchy $\chi_0<\chi_q<\chi_s$ that we have
measured above $T_c$.

\subsection{Screening and susceptibilities}

In Euclidean quantum field theories at $T=0$, transfer matrices in
all directions are isomorphic, and one, for example, can focus on the
time-direction transfer matrix--- ${\mathbf T}=\exp(-Ha)$ (here $H$
is the Hamiltonian and $a$ the lattice spacing). The eigenvalues of
$\mathbf T$ determine the hadron spectrum. The transfer matrix and,
hence, its eigenvalues do not change with temperature.  Specifically,
the symmetry of the transfer matrix remains the rotational $O(3)$
symmetry which is used to classify particle states at $T=0$.

At finite temperature or chemical potential all directions are not
equivalent. As a result, $\mathbf T$ is not the same as any spatial
direction transfer matrix. These have a totally different symmetry, $\cal
C$, that of the cylinder, as discussed extensively in the literature
\cite{bernd,saumen,d4h}. If the two lowest eigenvalues belonging to
the scalar of $\cal C$ become degenerate, then a phase transition
occurs. In general, the eigenvalues of this transfer matrix determine
screening masses in equilibrium. Breaking of $O(3)$ to $\cal C$ implies
strange phenomena like the `mixing' of different spins, or different
states of the same spin taking on independent dispersion relations
\cite{chin}. While these phenomena are strange, they are not new--- the
group theory is familiar from the textbook examples of the comparison
between the hydrogen atom and the hydrogen molecular ion $H_2^+$.

Lattice studies of screening correlators have hovered on the verge of this
phenomenology. If it has not received widespread attention in the past,
that is merely due to the practical difficulties of measuring some of
the non-trivial irreps of $\cal C$, as we show for the first time in
Table \ref{tb.masses}.  The equality of parity projected correlators
was used in Section \ref{sc.scr} to point out that the $\rho$ and
$\omega$ mix in screening, as do the $a_1$ and $b_1$.  The equality of
screening masses also gives some evidence for the mixing of $\pi$ with
the $J_z=0$ component of the $\rho/\omega$, causing a splitting in the
screening masses of different components of the latter.  We argued in
Section \ref{sc.chi} that the observed relation between $\chi_3$ and
the screening mass $M_S$ or between $\chi_{ud}$ and $M_\pi$ comes from
precisely the same physics as the mixing of different spins at finite
temperature. The same argument also implies that since fluctuations are
observable, screening phenomena are physical.

\subsection{Future directions}

Several directions for future work are clear, and have been discussed
in the body of the paper. Here we collect what seems to us the most
important and fruitful possibilities.

One direction for future work is to examine as many of the non-local
screening correlators as possible, in order to gather further information
on all the quantum numbers that screening correlators can come in
\cite{d4h}. As we have seen, this requires large lattices and immense
statistics, and may well profit from the use of new noise reduction
techniques and improved actions which do not change the symmetries
(or the positivity) of the transfer matrix.

The ratio $\chi_3/\chi_{FFT}$ is not explained in perturbation theory.
We noted this in a quenched computation \cite{qsus}, and have verified
it in the dynamical QCD computation here. The small difference
(roughly 3\%) between the quenched and dynamical results is also
significant. Explanation of these results stand as invitations to
those who resum the continuum high-temperature perturbation theory.
Future lattice computations will need to push towards the continuum
limit. Work in this direction is in progress, and will be reported soon.

As explained before, $\chi_{ud}$ vanishes in an ideal gas of quarks, but
in an interacting theory would take on a value of order $\alphas^3$. This
is several orders of magnitude larger than the largest result that our
measurements can tolerate. This observation also is an invitation to
perturbation theory.

The off-diagonal susceptibilities are non-vanishing only in the range
$0<T<T_c$. This region of temperature is hard to study, since all the
complications of $T=0$ QCD remain, and none of the simplifications of
$T>T_c$ QCD start. On the other hand it is important input to studies
of fluctuations in heavy ion collisions. We have reported a first
quantitative observation here. Future work will push towards more
realistic quark masses, taking the thermodynamic limit, and examining
a larger range of temperatures.

\end{document}